%% file: hel.tex
\documentstyle[sprocl,epsfig]{article}

\begin{document}

\include{abb}

\title{DIFFRACTION AT HERA    
    }

\author{P. MARAGE}

\address{Universit\'e Libre de Bruxelles -
    CP 230, Boulevard du Triomphe \\
    B-1050 Bruxelles, Belgium \\
    e-mail: {\tt pmarage@ulb.ac.be}}


\maketitle\abstracts{A review is presented of diffraction studies at HERA.
\footnote{
Talk given at the First Workshop on Forward Physics and Luminosity
  Determination at the LHC, Helsinki, Finland, November 2000.}
}


\section{Introduction: highlights of HERA}
						\label{sect:intro}

\subsection{The hard behaviour of the proton structure function $F_2$ at 
high energy}

With 27.5 GeV electrons or positrons colliding with 820 or 920 GeV 
protons through the exchange of highly virtual photons, HERA 
has been since 1992 an ideal machine for studying the proton structure 
at very high $\gamma^* p$ centre of mass energy $W$ 
(with $W^2 = y \cdot s$, $\sqrt {s}$ being the $e p$ centre of mass 
energy and $0 < y < 1$) or very low $x$ 
($x \simeq Q^2 / W^2$, \qsq\ being the negative square of the virtual 
photon four-momentum).

A major discovery at HERA has been the fast rise with energy of the 
$\gamma^* p$ cross section or, equivalently, of the $F_2$ structure 
function, in the deep inelastic scattering (DIS) domain.
This rise is larger for increasing \qsq:
when the cross section at low $x$ is parameterised as 
$\sigma (\gamma^* p) \propto x^{-\lambda}$, with $\lambda$ depending 
on $Q^2$, $\lambda \gsim 0.3$ for $Q^2 \gsim 100$ 
\gevsq~\cite{lambda} ($hard$ behaviour),
whereas $\lambda \simeq 0.08 - 0.10$ in hadron$-$hadron 
interactions~\cite{DoLa} ($soft$ behaviour).

\subsection{A large diffractive component in DIS}

Another major feature at HERA is the presence in the DIS domain 
of a large diffractive contribution, of about 8 \% of the total cross 
section. 

At high energy, diffractive and elastic scattering are governed by the 
exchange of the pomeron, an object carrying the vacuum quantum numbers.
Pomeron exchange also governs the high energy behaviour of total 
hadron cross sections, which are intimately related to elastic scattering
through the optical theorem.
The pomeron is thus an object of fundamental importance for particle 
physics, and it is a major challenge for QCD to provide a detailed 
description of diffractive scattering in terms of quark and gluon 
exchange.

At HERA, by changing the intermediate photon virtuality $Q^2$, it is 
possible to vary the resolution with which the pomeron structure is 
probed in diffractive interactions, and to study its partonic content. 
HERA has thus also become a wonderful tool for studying the QCD structure 
of diffraction, both providing a very rich amount of experimental results 
and triggering intense theoretical developments.


\section{Inclusive diffraction}
						\label{sect:inclusive}

\subsection{Kinematics}

\begin{figure}[htbp]
\vspace{-0.2cm}
\begin{center}
\epsfig{file=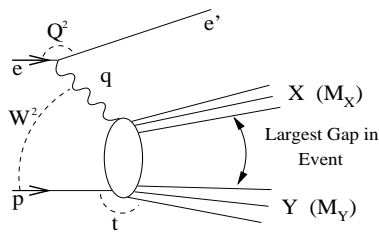,width=3cm,height=5cm,angle=270}
\end{center}
\vspace{-0.2cm}
\caption{Deep-inelastic diffractive interaction.}
\label{fig:rapiditygap}
\vspace{-0.2cm}
\end{figure}

The characteristic feature of diffraction
(see Fig.~\ref{fig:rapiditygap}) is that the final state hadronic system
is divided into two subsystems, ${X}$ and ${Y}$, 
separated by a large gap in rapidity devoid of hadronic energy.
The presence of the gap, due to the exchange of a colourless object, 
is attributed at high energy to pomeron exchange. 
In QCD, the simplest model for the pomeron is a pair of gluons.

When the proton remains intact (${Y} = p$, ``elastic'' scattering), 
the diffractive process
is defined, up to an azimuthal angle, by four kinematical variables:
$Q^2$, $\xpom$, $\beta$ and $t$, where $t$ is the squared
four-momentum transfer to the proton, and $x_{I\!\!P}$ and $\beta$  
are defined as
\begin{equation}
   x_\pom = 1 - x_L \simeq \frac{Q^2 + M_X^2}{Q^2 + W^2 } , \ \ \ \ 
    \beta \simeq \frac{Q^2 }{Q^2 + M_X^2} , \ \ \ \
    x = \beta \cdot x_\pom ,
                                                \label{eq:defs}
\end{equation}
with $x_L$ the fraction of the incident proton energy carried by the
scattered proton;
$\xpom$ is the fraction of the proton momentum 
carried by the exchange and $\beta$ is the fraction of the exchange 
momentum carried by the quark struck by the photon. 
Kinematics imply that a gap in rapidity is created 
between the system $X$ and the scattered proton when 
$x_\pom \ll 1$, i.e. $M_X \ll W$ and $x_L \simeq 1$.

Experimentally, diffractive events are thus selected by the direct 
observation of a large rapidity gap in the detector~\cite{H194} or, 
equivalently, by using the $M_X$ distribution~\cite{ZEUSMX,ZEUSBPC} and 
exploiting the fact that diffractive interactions are characterised 
by a non-exponentially suppressed rapidity gap. 
In both cases, a remaining background of events with proton dissociation 
has to be statistically subtracted, since low mass ${Y}$ systems give no 
signal in the detectors and can not be separated from elastic proton 
scattering.

Diffractive events can also be selected using proton spectrometers 
which tag the scattered proton~\cite{f2d-ZEUSLPS}. 
This provides a clean measurement of elastic diffraction without proton
dissociation background, and allows a 
measurement of the $t$ distribution, but the acceptance is low, especially 
for low \xp, and the statistics accumulated so far are poor.

\subsection{Diffractive structure functions}
						\label{sect:f2d}

In analogy with non-diffractive DIS, the inclusive 
diffractive cross section is expressed in the form of a three-fold 
structure function (four-fold when $t$ is measured):
\begin{equation}
  \frac { {\rm d}^3 \sigma \ (e + p \rightarrow e + X + p) }
 { {\rm d}Q^2 \ {\rm d}x_{I\!\!P} \ {\rm d}\beta }
        = \frac {4 \pi \alpha^2} {\beta Q^4}
            \ (1 - y + \frac {y^2} {2 (1 + R_D) } )
            \ F_2^{D(3)} (Q^2, x_{I\!\!P} , \beta ) ,
                                            \label{eq:fdthreefull}
\end{equation}
where \RD\ is the ratio of the longitudinal and transverse diffractive 
cross sections, which has not been measured so far.

In the case of ``$Regge \ factorisation$''~\cite{IS}, \fdthree\ can be 
factorised in the form
\begin{equation}
F_2^{D(3)} (Q^2, x_{I\!\!P} , \beta ) =
     \Phi (x_{I\!\!P} ) \cdot F_2^D (Q^2, \beta) ,
                                               \label{eq:factoris}
\end{equation}
where $\Phi(x_\pom )$ can be interpreted as an effective flux. 
At high energy ($\xpom < 0.01$), pomeron exchange dominates whereas for
lower energy (higher $\xpom$), reggeon ($\rho$, $\omega$, $f$ mesons) 
exchange provides an additional significant contribution.

For fixed \xp, $F_2^D (Q^2, \beta)$ describes the universal partonic 
structure of the exchange ($DIS\ factorisation$) ~\cite{collins}, 
$\beta$ playing the role of $x$ for hadron structure.

\begin{figure}[htbp]
\vspace{-0.5cm}
\begin{center}
 \epsfig{file=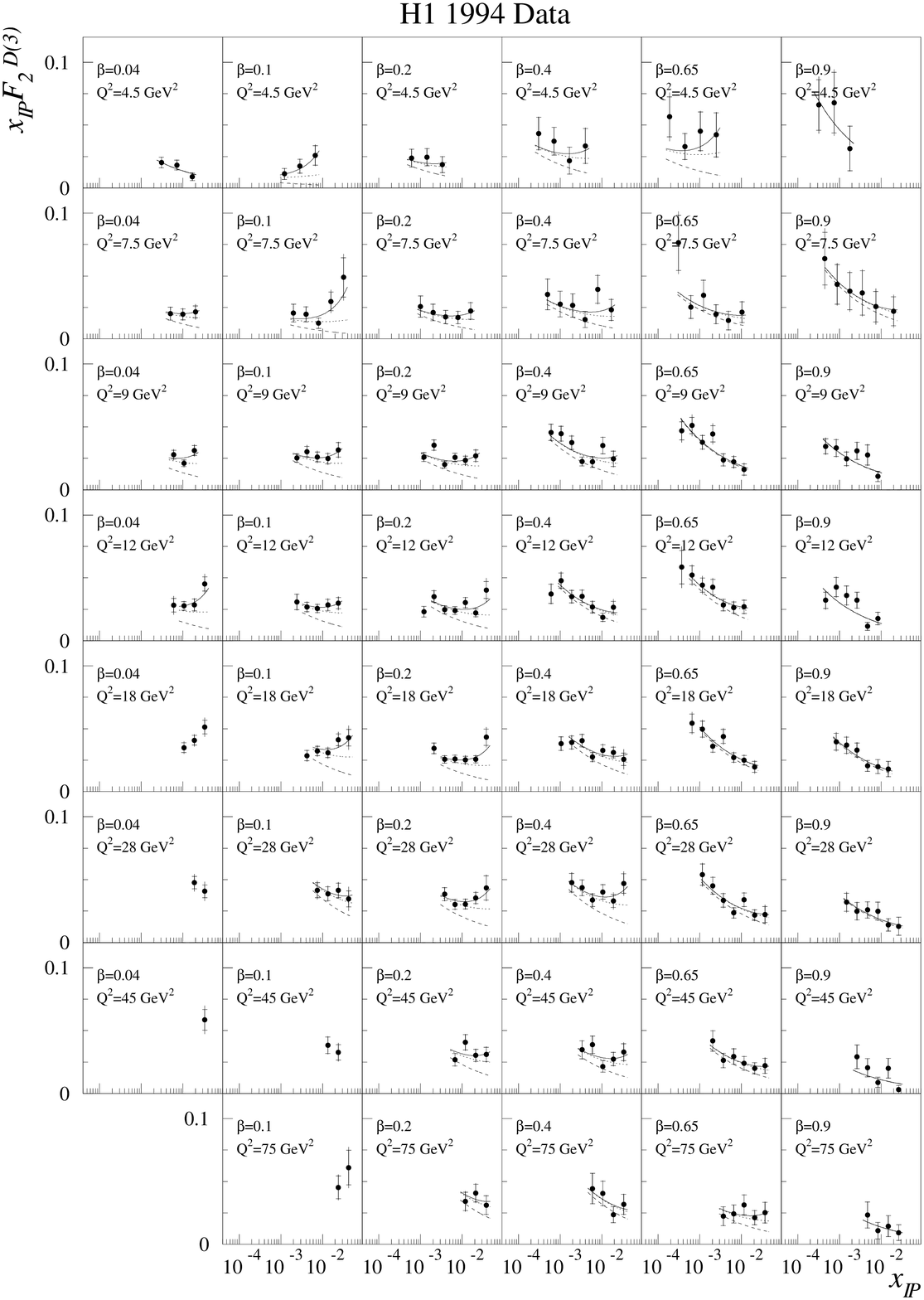,width=12cm,height=15cm}
\end{center}
\vspace{-0.5cm}
\caption{Measurement of 
$x_\pom\ \cdot$ \fdthreef\ ($M_Y < 1.6$ \gev ,
$|t| < 1 $ \gevsq ) as a function of $x_\pom$
for various \qsq\ and $\beta$ values~\protect\cite{H194}.
The curves show the results of the Regge fit with interference.
The dashed curves show the contributions of the pomeron alone, 
the dotted curves, the pomeron plus interference, and the continuous 
curves, the total.}
\label{fig:H1_f2d_94}
\vspace{-0.3cm}
\end{figure}

Fig.~\ref{fig:H1_f2d_94} presents the measurement~\cite{H194} of 
$\xpom \cdot F_2^{D(3)}$ as a function of $\xpom$ for several bins 
in \qsq\ and $\beta$.
It was fitted as the sum of a pomeron and a reggeon contribution, 
with possible interference~\footnote{
Only the $f$ meson is expected to
interfere with the pomeron.}, the fluxes being parameterised in 
a Regge inspired form:
\begin{equation}
  \Phi_{\Regg,\Pom}(\xpom ) \propto \xpom ^{n_{\Regg,\Pom}} , 
 \ \ \ \
  n_{\Regg,\Pom} = 2 \cdot \langle \alpha_{\Regg,\Pom} (t) \rangle - 1 .
                                                        \label{eq:flux}
\end{equation}
%
%
An exponential $t$ dependence, consistent with the 
data~\cite{tslope-ZEUSLPS}, is assumed.

In agreement with expectations, $\alpha_\regg (0)$ is found to be 
$0.50 \pm 0.18$, making a significant contribution typically for 
$\xpom \gsim 0.01$.
With the present statistics, the fit is not sensitive to the presence 
of a possible interference term.

\section{Energy dependence of diffraction; soft-hard interplay}
						\label{sect:energy}

\begin{figure}[htbp]
\vspace{-0.3cm}
\begin{center}
\epsfig{file=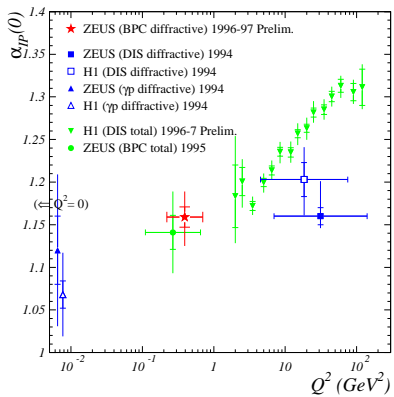,width=5.2cm,height=5.2cm}
\end{center}
\vspace{-0.3cm}
\caption{Measurements of $\alpha_\pom(0)$ 
 for diffractive and non-diffractive $ep$ interactions~\protect\cite{paul}.} 
\label{fig:alphaz}
\vspace{-0.3cm}
\end{figure}

The energy dependence of diffraction, measured by \apomz\ for fixed \qsq,
is shown in Fig.~\ref{fig:alphaz}.
In photoproduction~\cite{photoprod-h1,photoprod-zeus}, the value of 
\apomz\ is consistent with the
hadron$-$hadron case, as expected in view of the hadronic nature of real
photons.
In contrast, the measurements of \apomz\ for $Q^2 \neq 0$
are significantly higher~\cite{H194,ZEUSMX,ZEUSBPC}, indicating 
that naive Regge factorisation~(\ref{eq:factoris},\ref{eq:flux}) 
is broken, since \apomz\ depends on \qsq.

In a QCD approach, where diffraction is attributed to parton exchange, 
a \qsq\ dependence of \apomz\ is no surprise, since parton densities at 
low $x$ rise faster at higher \qsq, as measured by $F_2$ (see also 
Fig.~\ref{fig:alphaz}).
However, for a given \qsq\ value, \apomz\ is lower in diffraction than 
for inclusive DIS.

This ``semi-hard'' behaviour of diffraction can be most easily understood 
when the process is discussed in the proton rest frame~\footnote{
It can be helpful when discussing diffraction to visualise 
the process either from the Breit frame, with emphasis on the  
pomeron partonic content (structure function approach), or from the 
proton rest frame, which insists on the hadronic fluctuations of the 
photon (dipole approach~\cite{amirim}). 
These complementary pictures have of course to reconcile when physical 
measurements are discussed.}.
In this frame, because of the large boost, the photon has time to 
fluctuate, far from the target, into definite hadronic states which 
interact diffractively with the proton:
$|\gamma \rangle = |q \bar q \rangle + |q \bar q g \rangle + ...$

Two possible topologies can be contrasted:

a. The photon fluctuates into a large $k_T$, small transverse size 
dipole~\footnote{
For $q \bar q$ fluctuations, the dipole is formed by the two quarks;
in the case of $|q \bar q g \rangle $ fluctuations, colour factor 
considerations favour the topology consisting of a $(q \bar q)$ 
pair opposed to the gluon.}, with transverse dimension 
$r_{\perp} \propto 1/ Q^2$.
This large $k_T$ topology, which implies hard scattering (as for $F_2$), 
is kinematically preferred.
However, the small size of the colour dipole induces mutual screening: 
seen from the proton, the system appears as nearly colour neutral, 
and the interaction cross section is thus strongly reduced 
(this phenomenon is known as $colour \ transparency$).

b. The photon fluctuates into a longitudinally asymmetric, 
small $k_T$, large transverse size dipole (``aligned jet model''). 
As for hadron$-$hadron interactions, the cross section is large and, 
in the absence of a hard scale, the energy behaviour is soft, but photon 
fluctuations into this topology are kinematically disfavoured.

In contrast with the total cross section at high \qsq, 
which exhibits a purely hard behaviour, inclusive diffraction is thus a
semi-hard process, which includes both a hard component (small size 
dipoles, kinematically preferred but damped by colour transparency) 
and a soft component (large size dipoles, with large cross sections but 
small fluctuation probabilities).

The semi-hard nature of inclusive diffraction is confirmed by the
measurement of the exponentially falling $t$ distribution:
${\rm d} \sigma / {\rm d} t \propto exp (b \cdot t)$.
The ZEUS LPS measurement~\cite{tslope-ZEUSLPS} is
$b = 6.8 \pm 0.9 \pm 11 \ $ \gevsqm, smaller than for soft
processes ($b \simeq 10-12 \ $ \gevsqm\ for $\rho$ meson 
photoproduction~\cite{rho-photoprod-h1,rho-photoprod-zeus})
but larger than for a typical hard process
($b \simeq 4 - 4.5 \ $ \gevsqm\ for \jpsi\ 
production~\cite{psi-h1,psi-zeus}).

\section{Parton distributions; higher twists}
						\label{sect:pdf}

In the Breit frame, the \fdtwo\ structure function obtained from
\fdthree\ for a fixed value of \xp\ describes the partonic content of the 
exchange.
As for $F_2$, this structure function follows the DGLAP evolution
equations, and a universal set of parton distributions 
can be extracted from scaling violations~\footnote{
It must be stressed that DIS factorisation applies for fixed \xp\ and for
any mix of pomeron and reggeon in the exchange; Regge 
factorisation~(\ref{eq:factoris},\ref{eq:flux}) does not have to hold.}.

\begin{figure}[htbp]
\vspace{-0.cm}
\begin{center}
 \epsfig{file=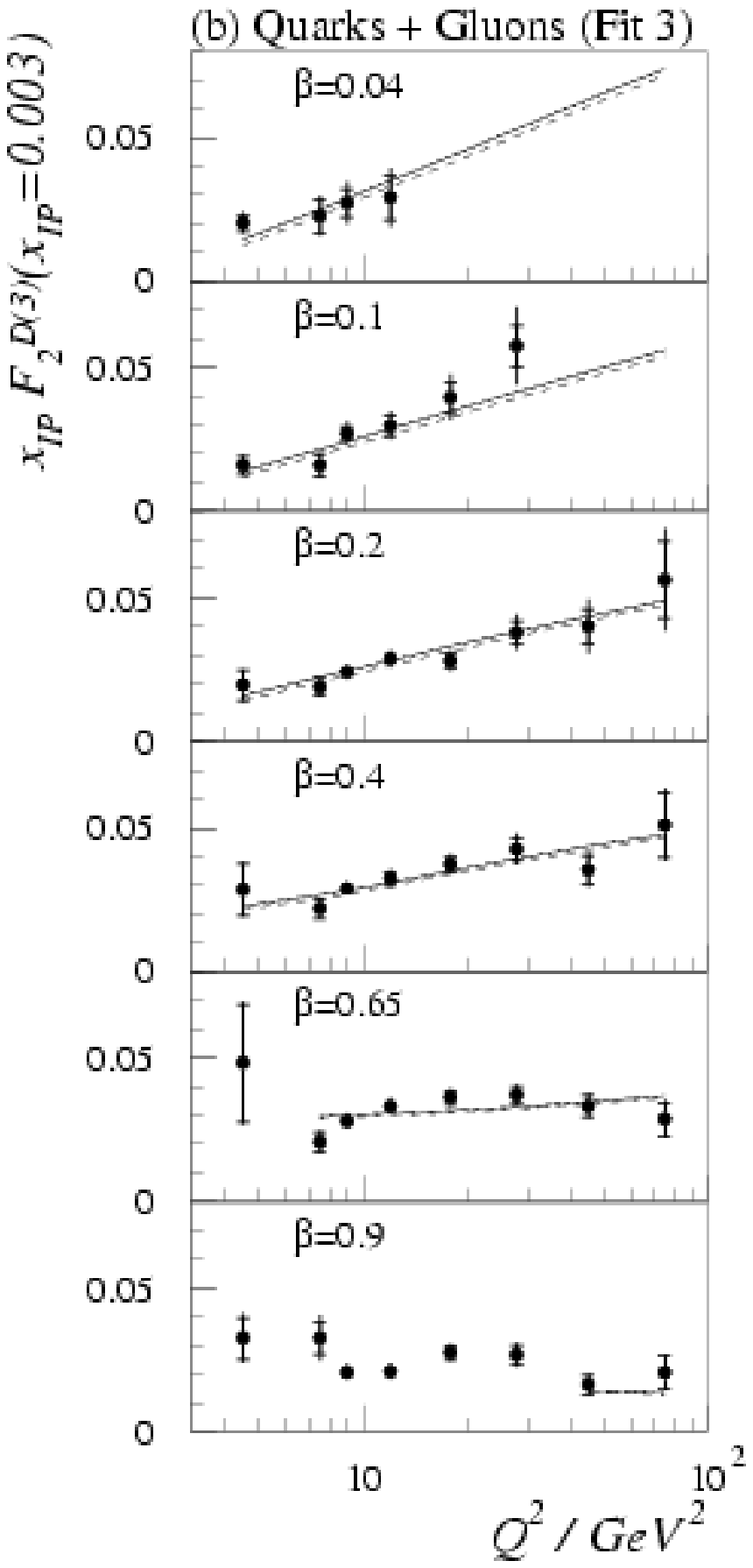,width=4.8cm,height=6.5cm}
 \epsfig{file=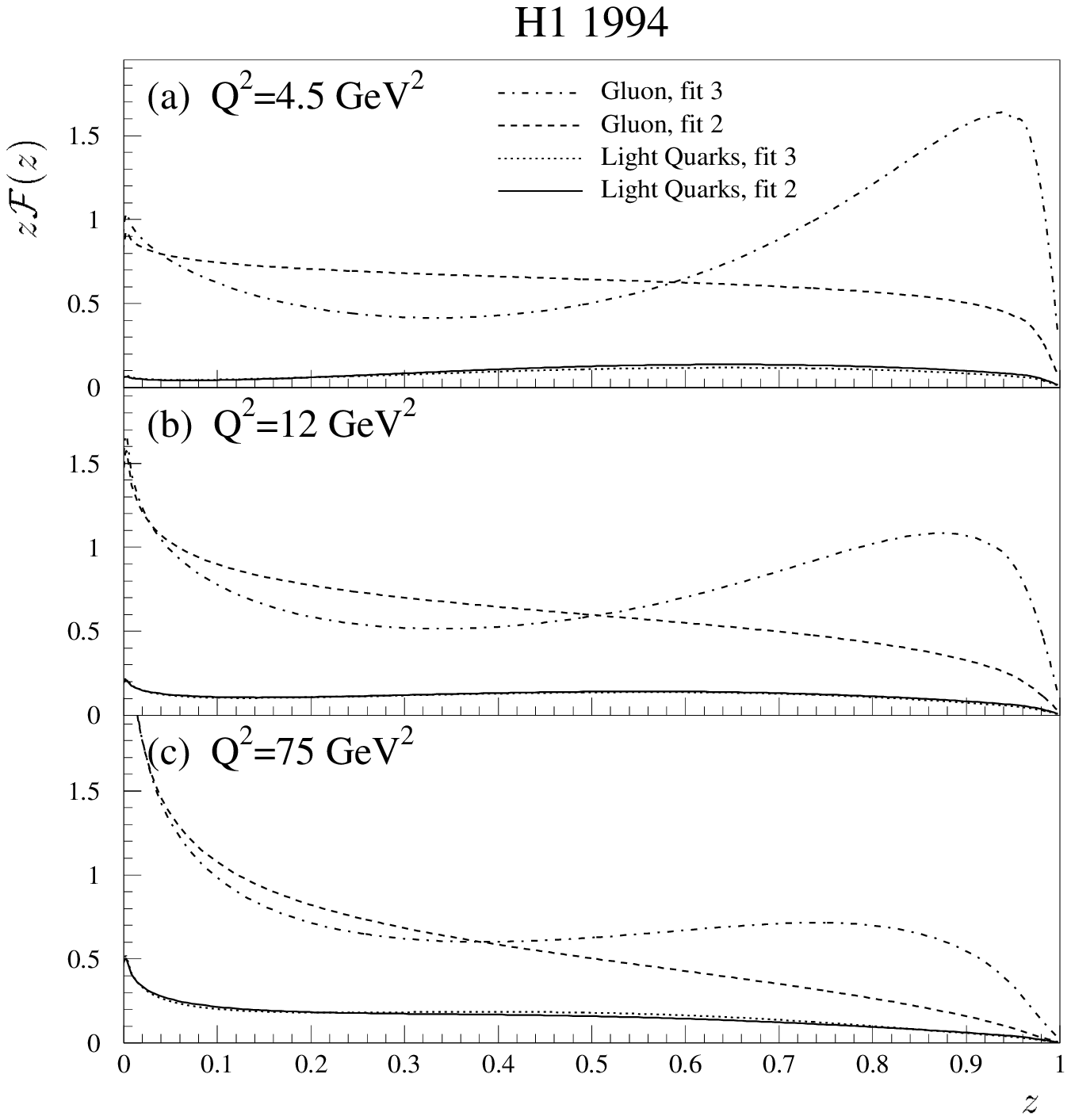,width=5.5cm,height=6.cm}
\end{center}
\vspace{-0.3cm}
\caption{Left: H1 measurement of the structure function 
 $\xpom \cdot \fiidiii$ for $\xpom = 0.003$ as a function of \qsq\ 
 in bins of $\beta$~\protect\cite{H194}; the curves are the result of 
 a QCD fit.
 Right: parton distributions in the pomeron.}
\label{fig:bis}
\vspace{-0.5cm}
\end{figure}

At a \xp\ value where the reggeon contribution is negligible compared
to the pomeron, Fig.~\ref{fig:bis} (left) shows that scaling violations 
are positive even at large $\beta$, in contrast
with hadron structure functions, which decrease for 
increasing \qsq\ at large $x$. 
This suggests the presence of a large gluon component at 
large $\beta$ in the pomeron, at variance with a very small gluon content
at large $x$ for hadrons.

Parton distributions in the pomeron, obtained from QCD fits, are
shown in Fig.~\ref{fig:bis} (right).
Gluons carry some 80\% of the pomeron momentum, and they dominate over 
quarks in the full $\beta$ range~\footnote{
The difference between the two acceptable fits (``flat'' and ``peaked''
gluons in Fig.~\ref{fig:bis}~right) affects mostly the region with
$\beta > 0.7$, which is discarded from the fit because it is affected by
large higher twist effects.}.

A specific feature of diffraction is the presence at high $\beta$ 
of a large higher twist component, which persists for large \qsq\ 
values.
Calculations in the proton rest frame, where the pomeron is 
modelled as a two gluon system~\cite{bekw}, indicate that three 
contributions dominate diffractive interactions: 
leading twist $q \bar q$ and $q \bar q g$ contributions from transverse 
photons, respectively in the intermediate $\beta$ region and in 
the low $\beta$ (large diffractive mass) region, and a higher twist 
$q \bar q$ contribution from longitudinal photons, dominant at large 
$\beta$~\footnote{
A specific channel with a large longitudinal higher twist component is 
vector meson production - see section~\ref{sect:vm}.}.
Within measurement precision, the inclusive data agree with 
these predictions~\cite{ZEUSMX}.

\section{Hadronic final states}
						\label{sect:hfs}

According to the DIS factorisation theorem~\cite{collins}, parton 
distributions extracted from QCD fits to inclusive diffraction
can be exported to hadronic final states in DIS~\footnote{
It is important to note that factorisation only applies
for DIS processes, and is broken for hadron$-$hadron interactions and 
for resolved photon interactions.
In these cases, reinteractions between the coloured remnants during 
the interaction time can fill the rapidity gap, and the diffractive 
cross section is significantly reduced, as observed in particular 
at the Tevatron, when parton densities obtained at HERA are compared 
to diffractive dijet, $W$ or bottom production~\cite{cdf}.}.

\begin{figure}[tbp]
\vspace{-0.cm}
\begin{center}
\epsfig{file=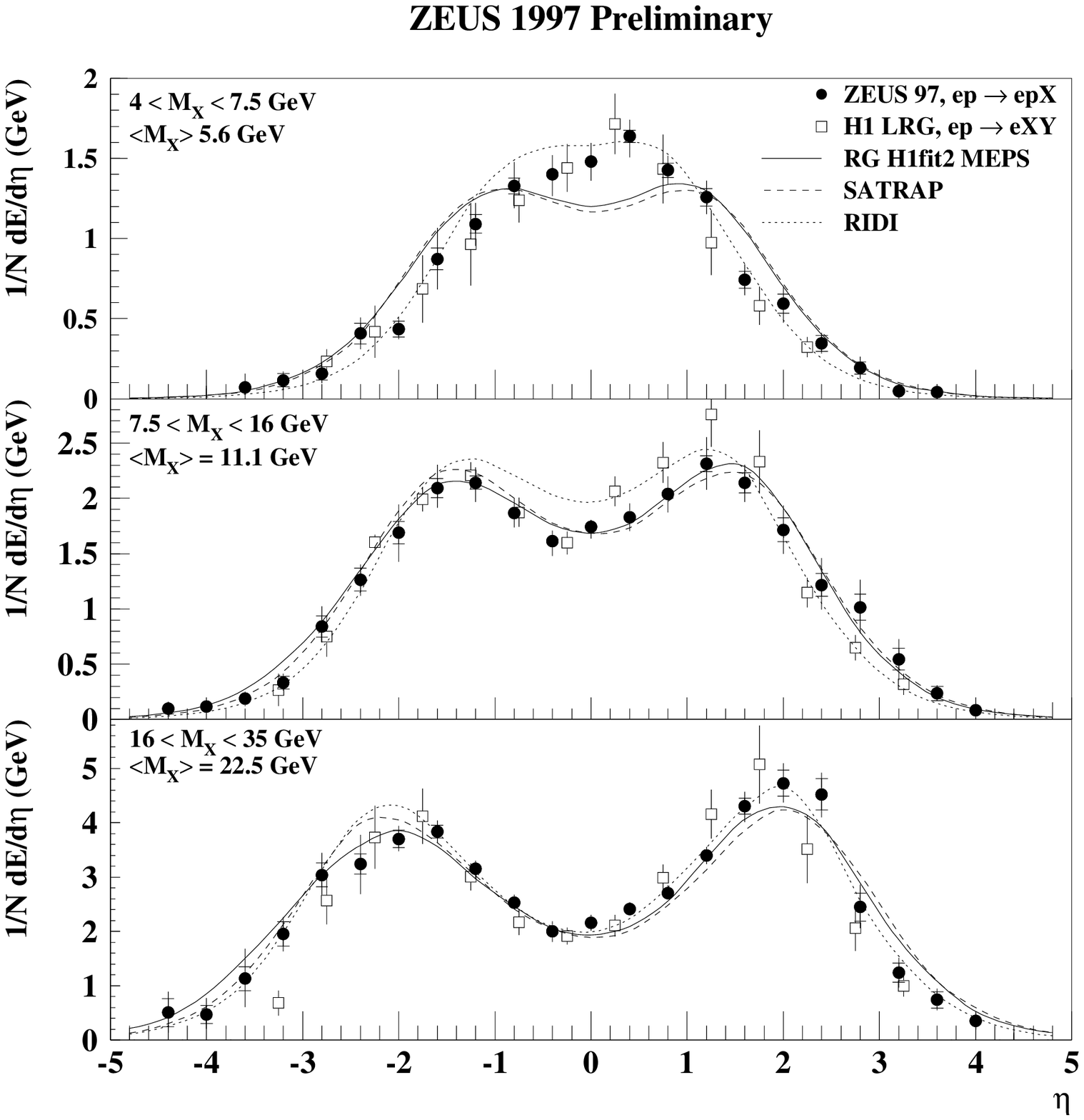,width=5.5cm,height=6cm}
\epsfig{file=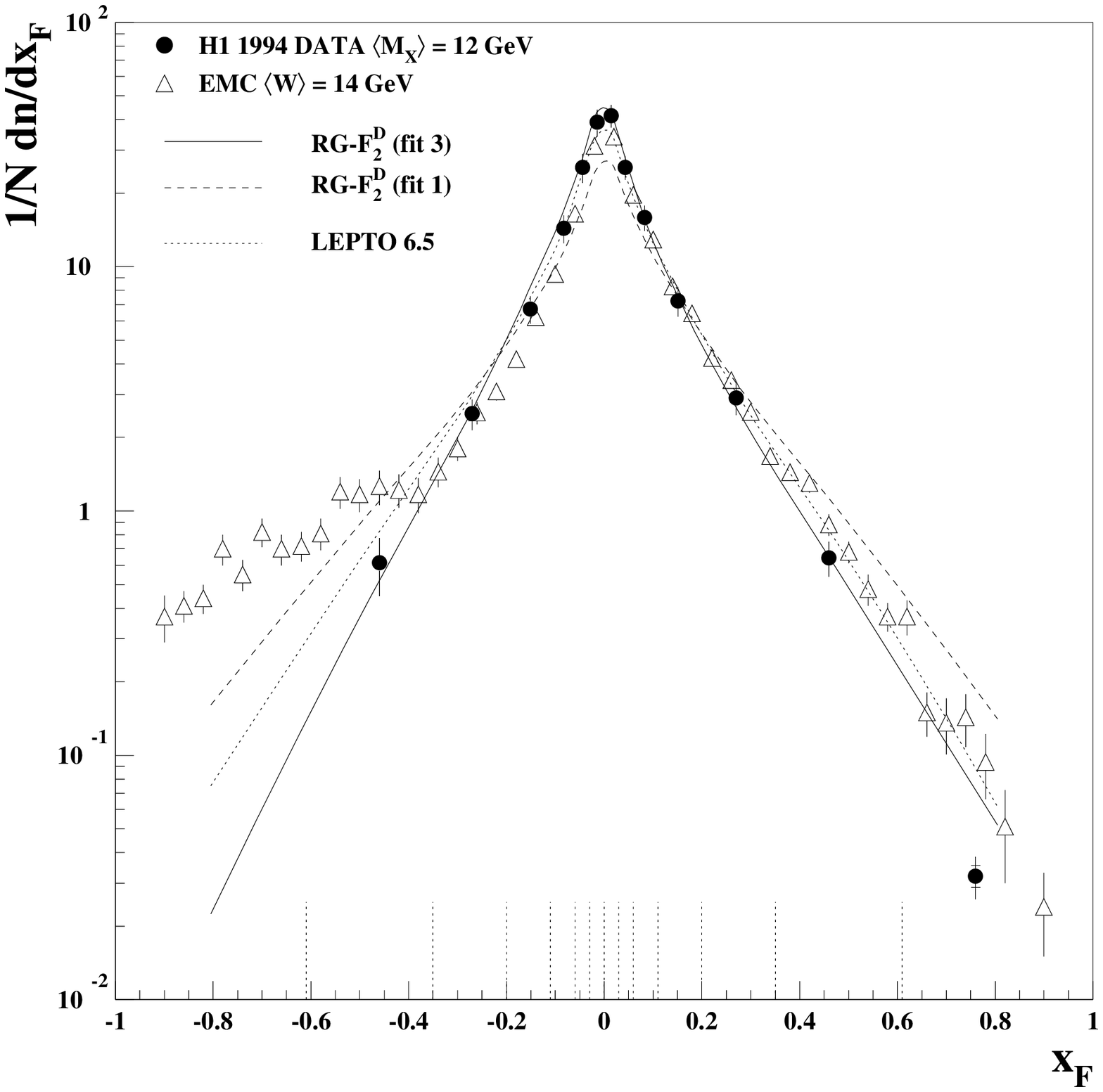,width=4.9cm,height=6.4cm}
\end{center}
\vspace{-0.5 cm}
\caption{Left: energy flow for diffractive 
 interactions~\protect\cite{zeus-final-state};
 right: $x_F$ distribution~\protect\cite{h1-final-state}; 
 the measurements are compared to Monte-Carlo
 predictions using parton distributions extracted from inclusive
 diffraction.}
\label{fig:hfs}
\vspace{-0.5cm}
\end{figure}

Semi-inclusive variables (energy flow, $x_F$, transverse momentum of tracks, 
multiplicity distributions) present marked differences with 
non-diffractive interactions, but similarities to $e^+ e^-$ annihilation.
These features are reasonably well described by Monte-Carlo simulations 
using parton distributions otained from inclusive 
diffraction~\cite{zeus-final-state,h1-final-state} 
$-$ see Fig.~\ref{fig:hfs}. 

Of particular interest is the measurement of diffractive 
dijet~\cite{h1-jets,zeus-jets}
and charm~\cite{zeus-charm,h1-charm}
production, since the implied boson gluon fusion process provides 
a direct probe of the gluon content of the pomeron. 

\begin{figure}[tbp]
\vspace{-0.cm}
\begin{center}
\epsfig{file=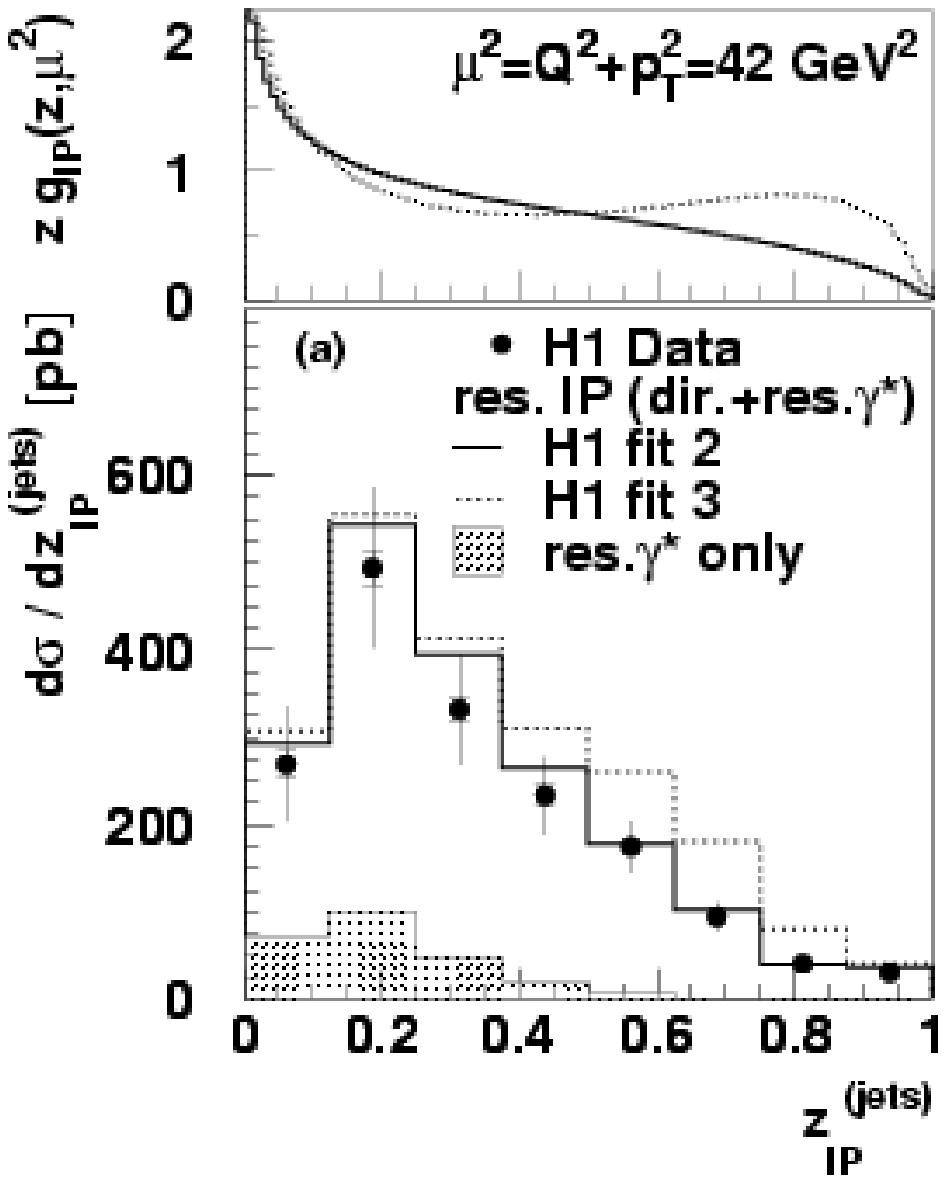,width=4.5cm,height=5.5cm}
\epsfig{file=whitebox.eps,width=0.6cm,height=5cm}
\epsfig{file=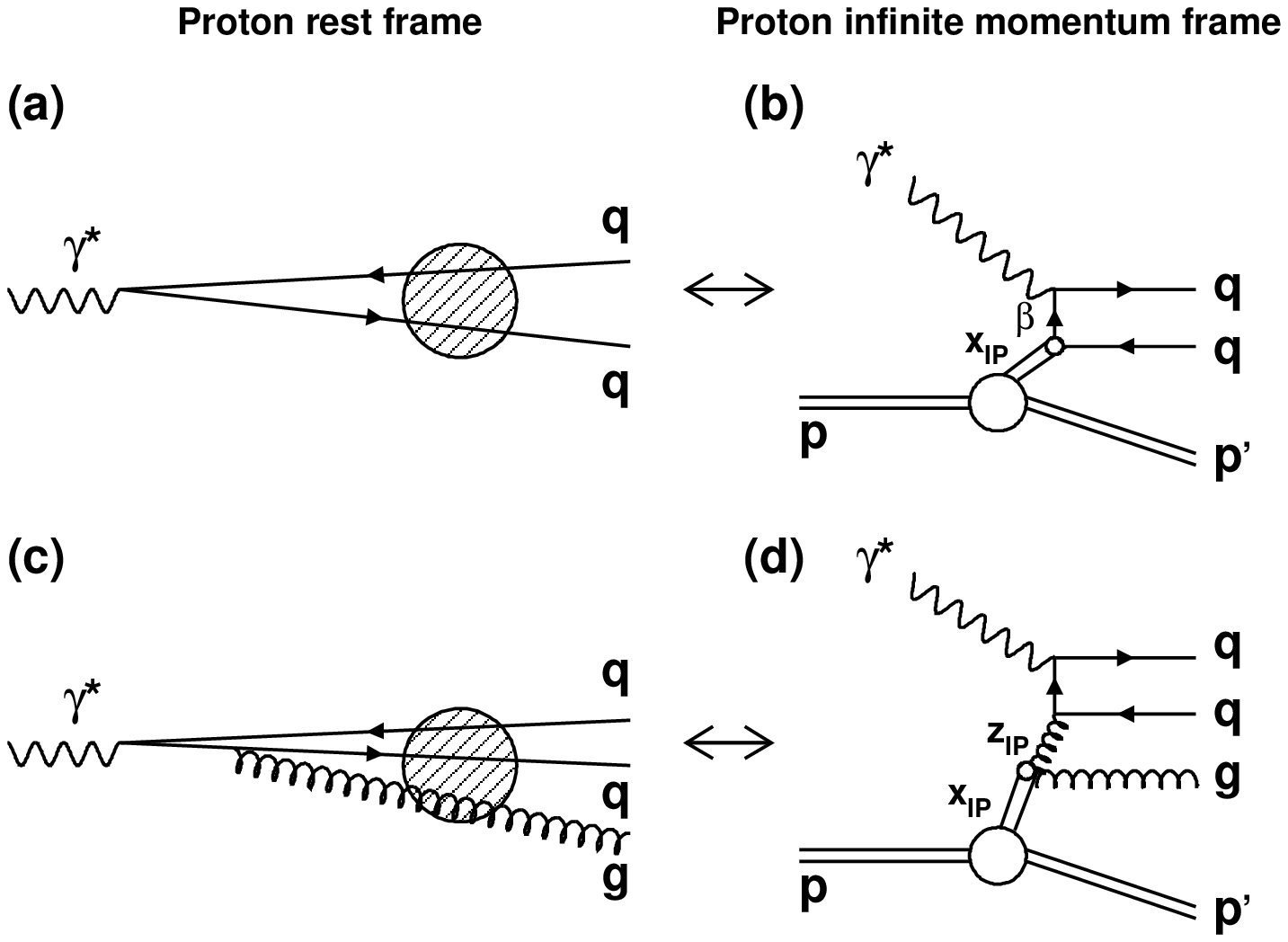,width=6.5cm,height=5cm}
\end{center}
\vspace{-0.3cm}
\caption{Left: diffractive dijet production~\protect\cite{h1-jets} as a
 function of $z_\Pom$, compared to Monte-Carlo predictions using
 respectively the H1 ``flat'' and ``peaked'' gluons obtained from QCD fits
 to \protect\fdthree; the contribution of direct and
 resolved photon contributions are indicated.
 Right: two complementary pictures of diffractive dijet production.} 
\label{fig:jets}
\vspace{-0.5cm}
\end{figure}

A remarkable description of kinematical variables in diffractive dijet 
electroproduction~\cite{h1-jets} is achieved by a Monte-Carlo simulation 
which includes the H1 ``flat'' gluon distribution (see 
Fig.~\ref{fig:jets}~left).
An interesting feature is the broad distribution of $z_\Pom$, the  
pomeron momentum fraction carried by the two jets, and the absence of a 
peak for $z_\Pom \simeq 1$.
In a Breit frame picture, this indicates the presence of pomeron remnants,
as expected for a gluon dominated pomeron
(see Fig.~\ref{fig:jets}~d).
In a proton rest frame approach, this diagram corresponds to
$|q \bar q g \rangle$ states with a low $p_T$ gluon 
(Fig.~\ref{fig:jets}~c). 
In contrast, the contribution of large $p_T$, small size 
$|q \bar q \rangle $ states (Figs.~\ref{fig:jets}~a) 
is damped by colour transparency.

For charm production, in spite of very limited statistics, similar 
features are observed.

\section{Exclusive vector particle production}
						\label{sect:vm}

The exclusive production of vector mesons and photons provides a
rich diffraction laboratory:
detailed studies are performed for different values of the scales 
provided respectively by the mass of the constituent quarks, \qsq\ 
and $t$.

\begin{figure}[tbp]
\vspace{0.4cm}
\begin{center}
 \epsfig{file=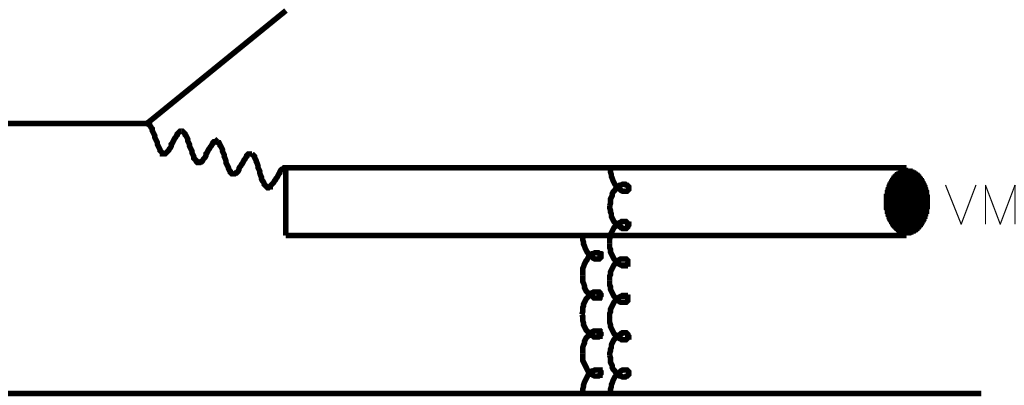,width=3.5cm,height=1.8cm}
 \epsfig{file=whitebox.eps,width=1.cm,height=1.8cm}
 \epsfig{file=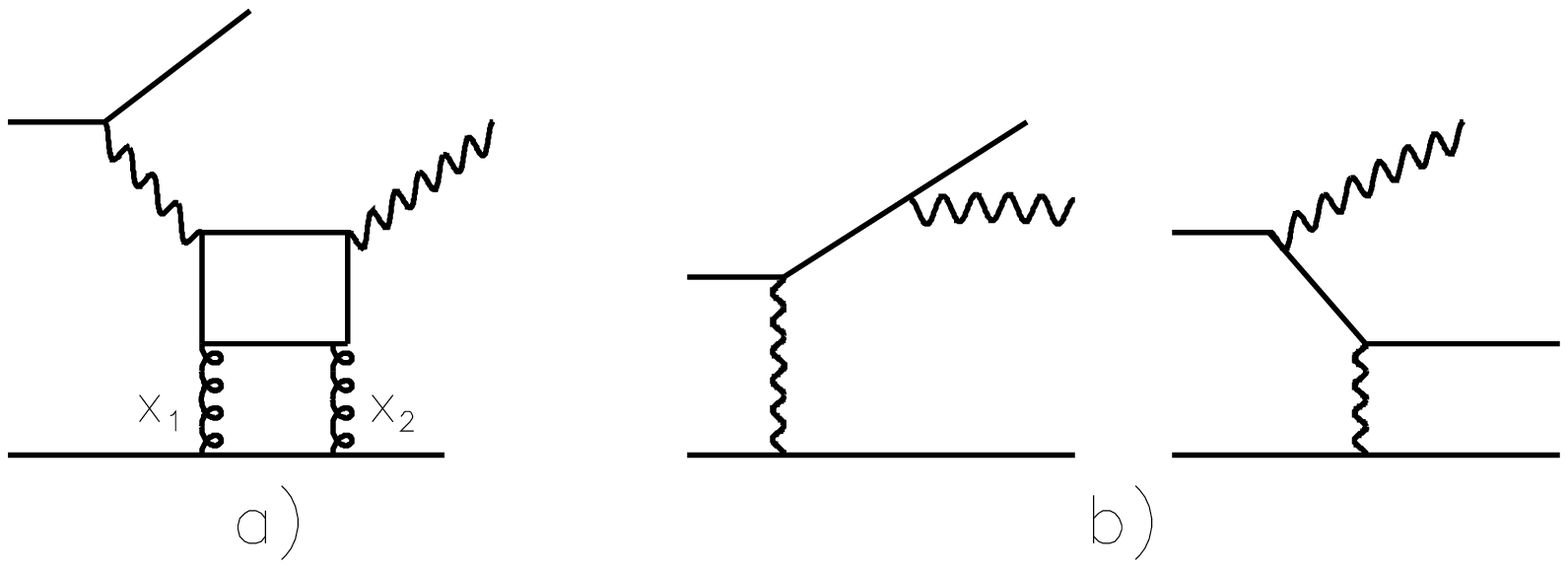,width=6.cm,height=1.8cm}
\end{center}
\vspace{-0.4cm}
\caption{Left: exclusive vector meson production;
 right: a) deeply virtual Compton scattering; 
 b) Bethe-Heitler (QED Compton) background to DVCS.} 
\label{fig:diag}
\vspace{-0.5cm}
\end{figure}

\begin{figure}[tbp]
\setlength{\unitlength}{1.0cm}
\vspace{-0.cm}
\begin{center}
 \epsfig{file=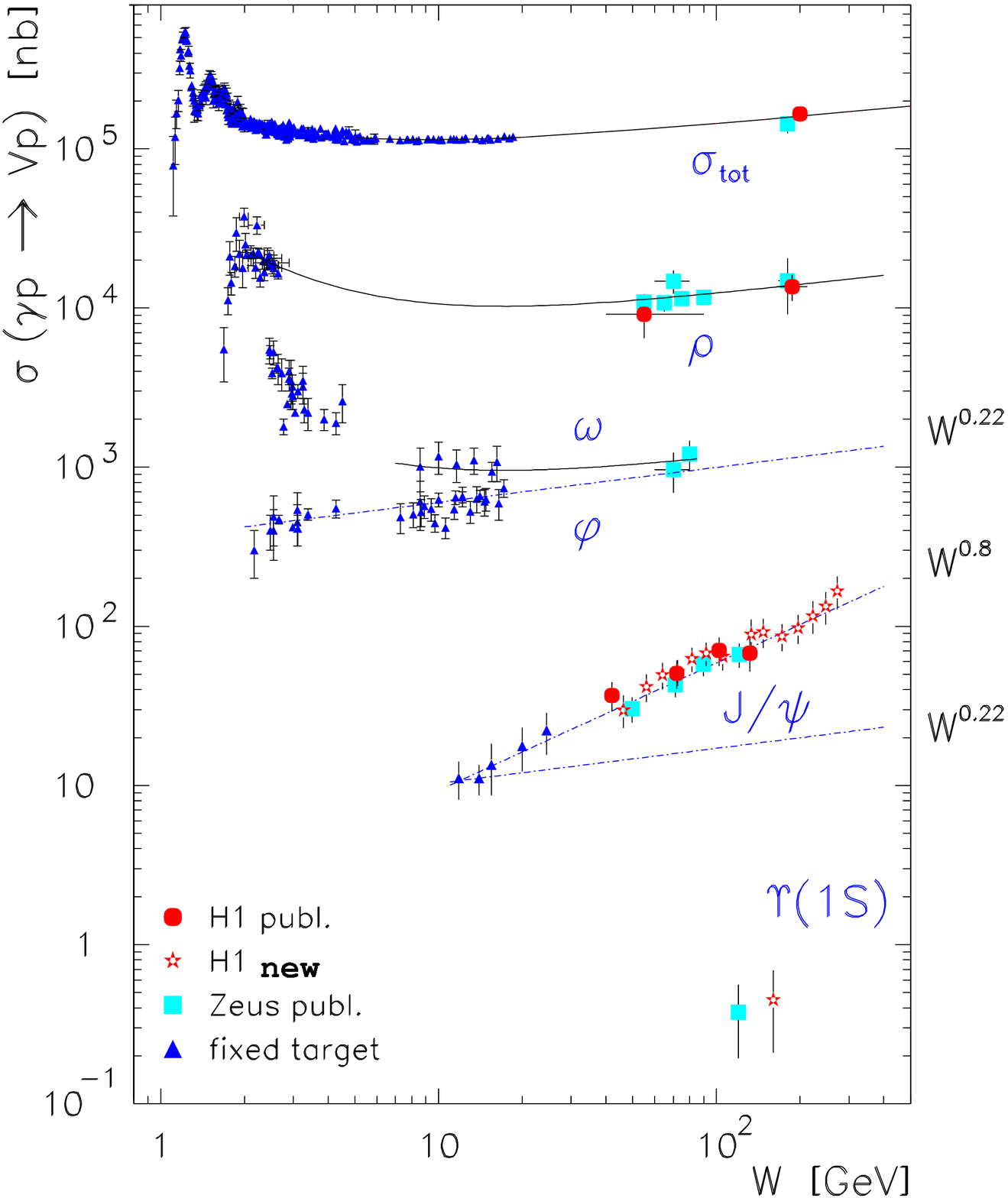,width=5.8cm,height=5.8cm}
\begin{picture}(5.8,5.8)
 \put(0.0,0.){\epsfig{file=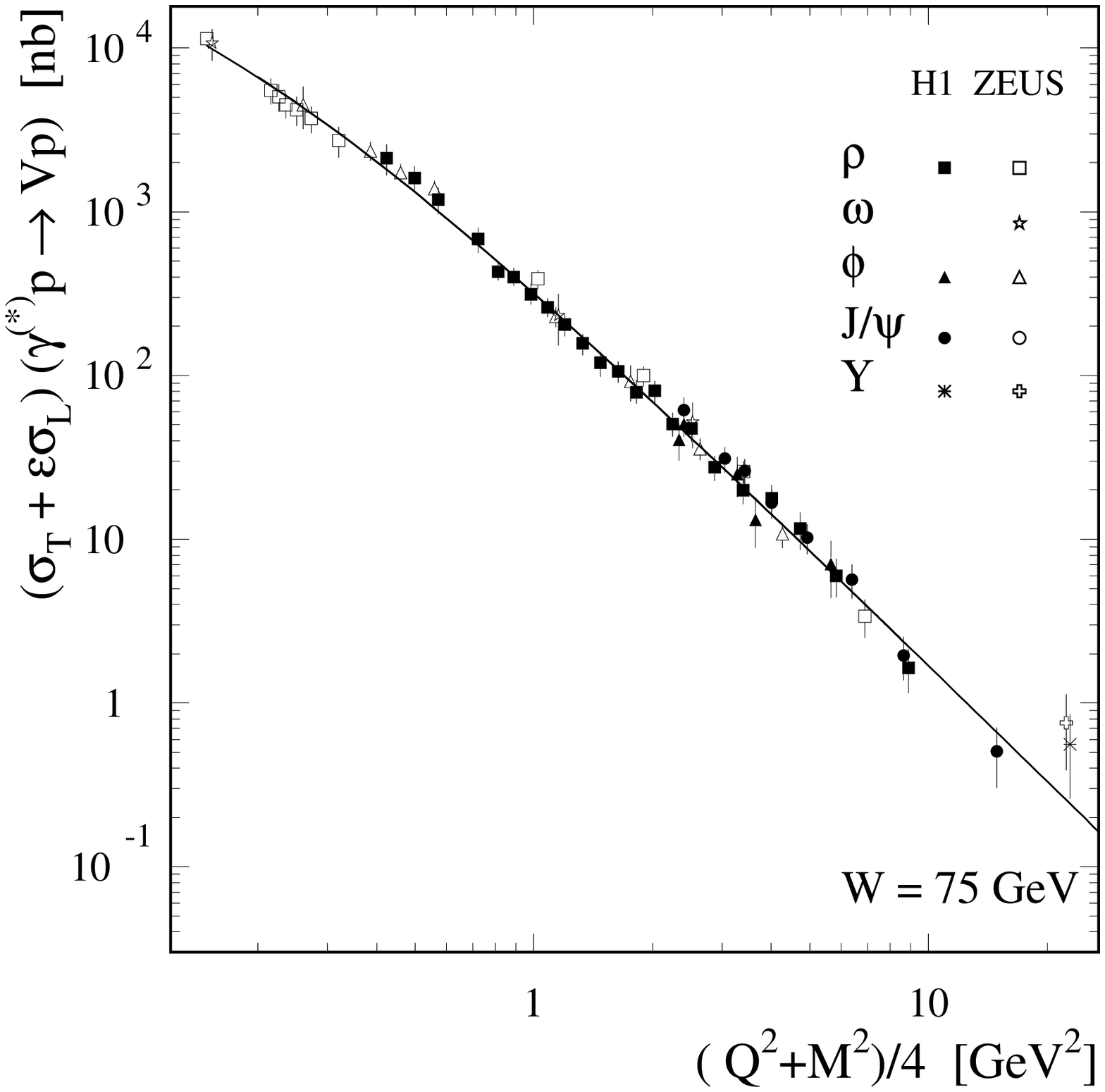,width=5.8cm,height=5.8cm}}
 \put(1.1,1.){\epsfig{file=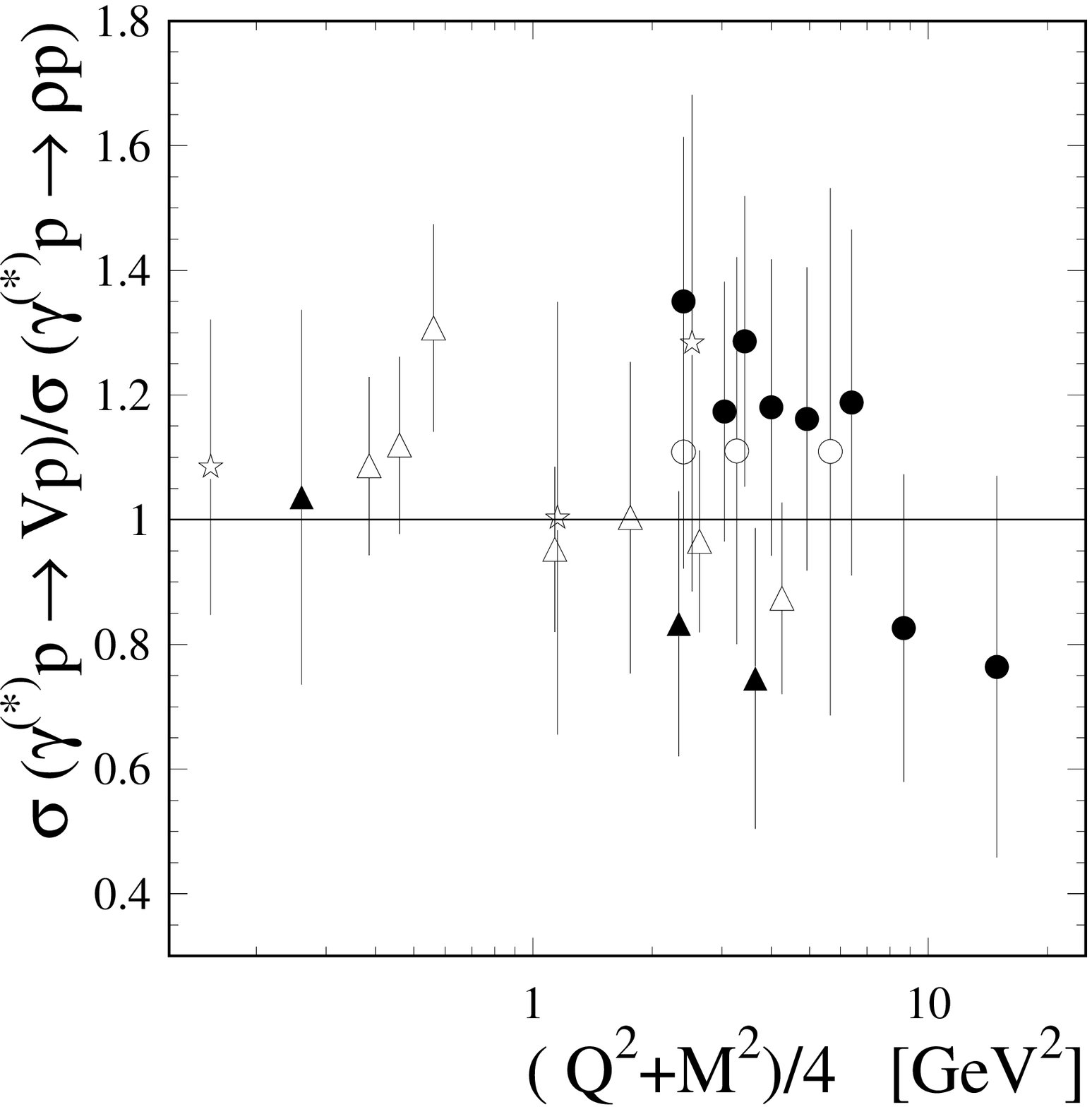,width=2.85cm,height=2.25cm}}
\end{picture}
\end{center}
\vspace{-0.3 cm}
\caption{Left: energy dependence for $\sigma_{tot} (ep)$ and for
 $\rho, \omega, \phi, J/\psi $ photoproduction;
 right: $Q^2 + M_V^2$ dependence for photo- and electroproduction of
 several vector mesons, scaled by the SU(4) 
 factors~\protect\cite{phi-h1}.} 
\label{fig:vm}
\vspace{-0.5cm}
\end{figure}

A striking effect is the hard energy dependence of \jpsi\ 
photoproduction~\cite{psi-h1,psi-zeus}
(\apomz\ $\gsim 0.25$), much stronger than for light 
vector meson photoproduction (see Fig.~\ref{fig:vm}~left).
This is an effect of the large charm mass, which implies that
the process takes place over short distances and probes directly the 
hard gluon content of the proton:
$\sigma (e\ p \rightarrow e\ p\ J/\psi ) \propto |x G(x)|^2$.
Similarly, the energy dependence of $\rho$ and $\phi$ 
electroproduction~\cite{rho-zeus-osaka,rho-h1,phi-h1}
increases with \qsq.

It is remarkable that the cross sections for different vector mesons, 
which differ by large factors in photoproduction, are very similar, 
up to their quark content (SU(4) factors), once plotted as a function 
of $Q^2 + M_V^2$$-$ see Fig.~\ref{fig:vm}~right.
This observation, in agreement with QCD expectations, confirms the role
of the quark mass and of \qsq\ as hard scales.

\begin{figure}[tbp]
\setlength{\unitlength}{1.0cm}
\vspace{-0.cm}
\begin{center}
 \epsfig{file=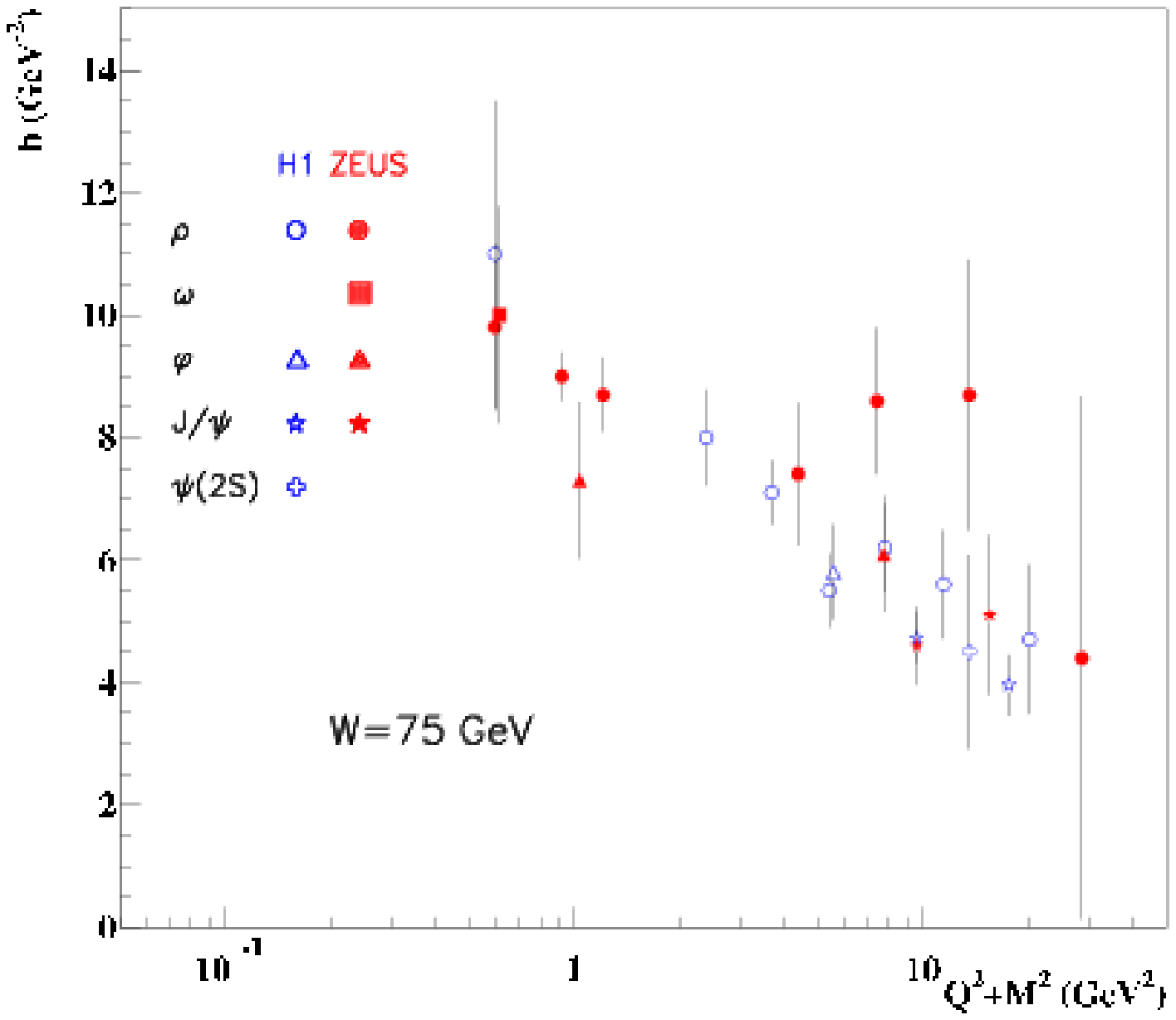,width=5.5cm,height=4.5cm}
 \epsfig{file=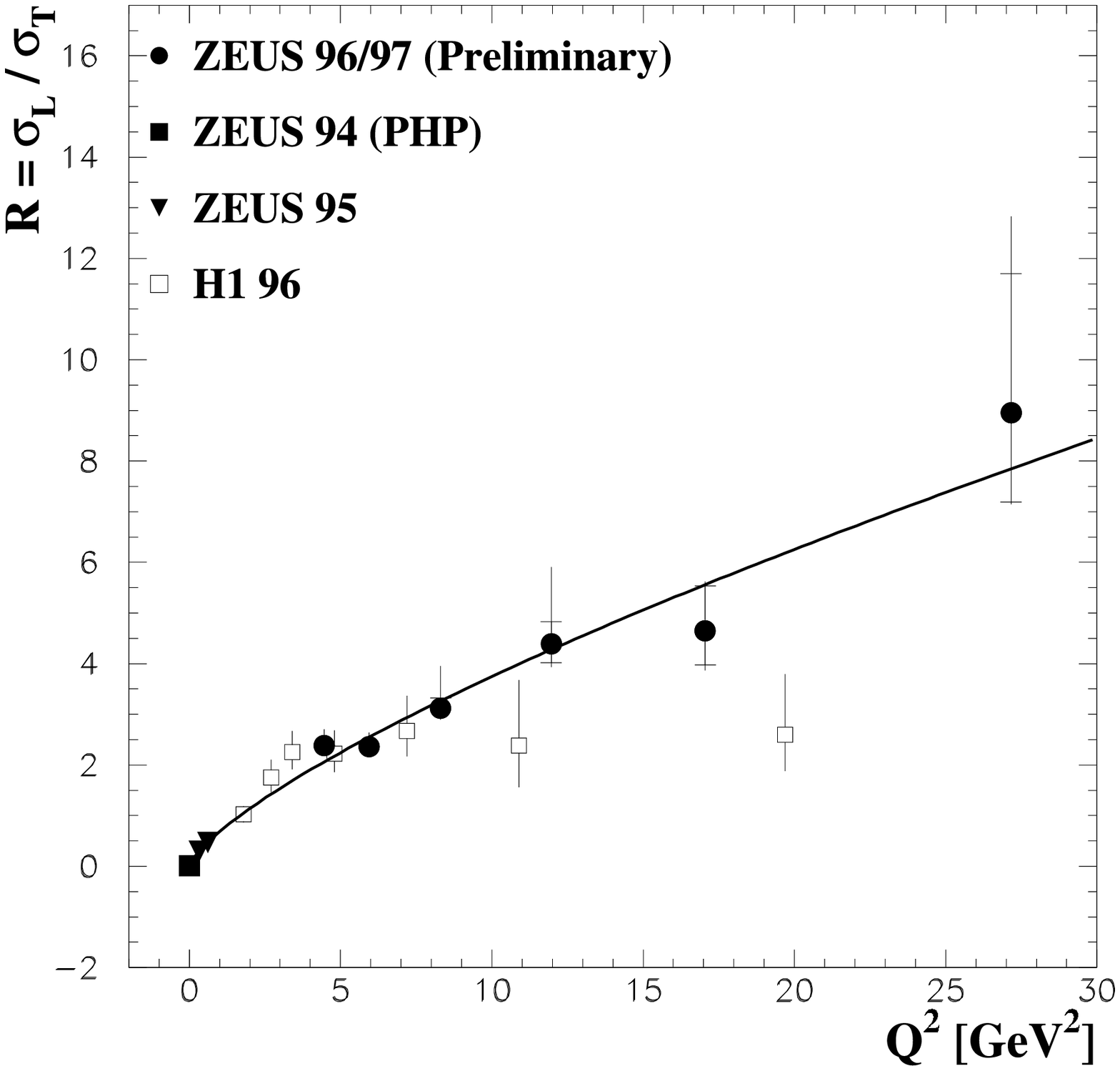,width=5.5cm,height=4.5cm}
\end{center}
\vspace{-0.3 cm}
\caption{Left: exponential $t$ slopes for several vector meson
 diffractive production, as a function of 
 $Q^2 + M_V^2$~\protect\cite{yoshida};
 right: $R = \sigma_L / \sigma_T$ as a function of \qsq\ for $\rho$
 meson electroproduction, the curve representing an empirical 
 fit~\protect\cite{rho-zeus-osaka}.} 
\label{fig:vm2}
\vspace{-0.5cm}
\end{figure}

\jpsi\ production is also characterised~\cite{psi-h1,psi-zeus}
by a value of the exponential
$t$ slope $\simeq 4 - 4.5 \ $ \gevsqm, much smaller than for light 
vector meson photoproduction. 
This is due to the small size of the \jpsi\ meson, the slope reflecting
the transverse size of the interacting objects. 
A decrease of the slope is also observed for increasing \qsq\ for 
$\rho$ and $\phi$ mesons, with an indication of the universality of the
slope as a function of $Q^2 + M_V^2$  
$-$ see Fig.~\ref{fig:vm2}~left.

The measurement of angular distributions for vector mesons
provides information on the helicity structure of diffraction. 
In particular, the ratio $R = \sigma_L / \sigma_T$ of the longitudinal to
transverse cross sections for $\rho$ electroproduction increases with
\qsq, as can be seen in Fig.~\ref{fig:vm2}~right.
A small but significant violation of $s$-channel helicity conservation is
observed~\cite{rho-h1,rho-zeus-mde}, the dominant helicity flip amplitude 
being from a transverse photon to a longitudinal meson; 
this is in agreement with QCD expectations~\cite{ivanov}.

The cross section for deeply virtual Compton scattering (DVCS), 
where the  virtual photon converts diffractively into a real photon 
as illustrated in Fig.~\ref{fig:diag}~right~a), has also been measured 
and found in good agreement with QCD predictions~\cite{dvcs-h1}.
This process allows the measurement in the proton of
skewed parton distributions, i.e. generalised parton distributions
including correlations between partons with different longitudinal
momenta.
This concept~\cite{spd} is introduced to describe the longitudinal 
momentum transfer kinematically necessary for
putting on mass shell the virtual photon or the vector meson. 
It is also important for the photoproduction of heavy vector mesons, 
in particular $\Upsilon$~\cite{psi-h1,zeus-upsilon}.

\section{Conclusions}

Numerous important results on diffraction have been obtained at HERA, 
which triggered new theoretical developments.

However some measurements, important to test models and achieve a deeper
QCD understanding of diffraction, are still missing, in particular 
measurements of the longitudinal cross section, of the dipole size 
in different processes (through measurements of the $t$ slope) and
of higher twist contributions. 
In addition, experimental uncertainties related to limited statistics 
in the presence of a hard scale (e.g. for charm production) 
and to the contamination of the elastic channel by proton dissociation 
background still affect the quality of the data.

Significant progress will be obtained from the large statistics 
accumulated recently and after the HERA luminosity upgrade.
The installation by H1 of a very forward proton spectrometer~\cite{pots} 
with full acceptance in $t$ for $\xpom \simeq 0.01$ will allow clean
selections of diffractive processes in the presence of a hard scale, 
provide measurements of the $t$ slopes for several processes, give 
information on the longitudinal cross section, and open a new field of
research through the comparison of elastic and proton dissociation events.


\section*{Acknowledgements}
It is a pleasure to thank the organisors of the Workshop, in
particular Risto Orava, for a very pleasant and fruitful meeting.



\end{document}

%% file: abb.tex
%
%
\newcommand{\Pom}{{I\!\!P}}
\newcommand{\Regg}{{I\!\!R}}
\newcommand{\xl} {$x_L$}
\newcommand{\ttt} {$t$}
\newcommand{\fdtwo} {$F_2^{D(2)}$}
\newcommand{\fdthree} {$F_2^{D(3)}$}
\newcommand{\fdthreef} {$F_2^{D(3)} (Q^2, x_{I\!\!P} , \beta )$}
\newcommand{\fdfour} {$F_2^{D(4)}$}
\newcommand{\fdfourf} {$F_2^{D(4)}  (Q^2, x_{I\!\!P} , \beta , t)$}
\newcommand{\fdc} {F_2^{D}_{charm}}
\newcommand{\flpom} {F_L^{\Pom}}
\newcommand{\RD} {$R_D$}
\newcommand{\xp}{$x_{I\!\!P}$}
\newcommand{\apomz}{$\alpha_{\Pom}(0)$}

\newcommand{\s}{\mbox{$s$}}
\newcommand{\ttra}{\mbox{$t$}}
\newcommand{\modt}{\mbox{$|t|$}}
\newcommand{\eminpz}{\mbox{$E-p_z$}}
\newcommand{\eminpzs}{\mbox{$\Sigma(E-p_z)$}}
\newcommand{\rap}{\ensuremath{\eta^*} }
\newcommand{\W}{\mbox{$W$}}
\newcommand{\w}{\mbox{$W$}}
\newcommand{\Q}{\mbox{$Q$}}
\newcommand{\q}{\mbox{$Q$}}
\newcommand{\xB}{\mbox{$x$}}  
\newcommand{\xF}{\mbox{$x_F$}}  
\newcommand{\xg}{\mbox{$x_g$}}  
\newcommand{\xbj}{$x$}
\newcommand{\xpom}{x_{I\!\!P}}
\newcommand{\zpom}{z_{I\!\!P}}
\newcommand{\y}{\mbox{$y~$}}
\newcommand{\Qsq}{\mbox{$Q^2$}}
\newcommand{\qzsq}{\mbox{$Q_o^2$}}
\newcommand{\qsq}{\mbox{$Q^2$}}
\newcommand{\kjet}{\mbox{$k_{T\rm{jet}}$}}
\newcommand{\xjet}{\mbox{$x_{\rm{jet}}$}}
\newcommand{\Ejet}{\mbox{$E_{\rm{jet}}$}}
\newcommand{\thjet}{\mbox{$\theta_{\rm{jet}}$}}
\newcommand{\pjet}{\mbox{$p_{T\rm{jet}}$}}
\newcommand{\et}{\mbox{$E_T$}}
\newcommand{\kt}{\mbox{$k_T$}}
\newcommand{\ptrans}{\mbox{$p_T~$}}
\newcommand{\pth}{\mbox{$p_T^h~$}}
\newcommand{\pte}{\mbox{$p_T^e~$}}
\newcommand{\ptsq}{\mbox{$p_T^{\star 2}~$}}
\newcommand{\as}{\mbox{$\alpha_s$}}
\newcommand{\ycut}{\mbox{$y_{\rm cut}~$}}
\newcommand{\gx}{\mbox{$g(x_g,Q^2)$~}}
\newcommand{\xpart}{\mbox{$x_{\rm part~}$}}
\newcommand{\mrsdm}{\mbox{${\rm MRSD}^-~$}}
\newcommand{\mrsdmp}{\mbox{${\rm MRSD}^{-'}~$}}
\newcommand{\mrsdn}{\mbox{${\rm MRSD}^0~$}}
\newcommand{\lambdams}{\mbox{$\Lambda_{\rm \bar{MS}}~$}}
%
%
\newcommand{\gp}{\ensuremath{\gamma}p }
\newcommand{\gammasp}{\ensuremath{\gamma}*p }
\newcommand{\gammap}{\ensuremath{\gamma}p }
\newcommand{\gsp}{\ensuremath{\gamma^*}p }
\newcommand{\dsiget}{\ensuremath{{\rm d}\sigma_{ep}/{\rm d}E_t^*} }
\newcommand{\dsigrap}{\ensuremath{{\rm d}\sigma_{ep}/{\rm d}\eta^*} }
\newcommand{\epem}{\mbox{$e^+e^-$}}
\newcommand{\ep}{\mbox{$ep~$}}
\newcommand{\epl}{\mbox{$e^{+}$}}
\newcommand{\emi}{\mbox{$e^{-}$}}
\newcommand{\epm}{\mbox{$e^{\pm}$}}
\newcommand{\xsec}{cross section}
\newcommand{\xsecs}{cross sections}
\newcommand{\inter}{interaction}
\newcommand{\inters}{interactions}
\newcommand{\eplp}{\mbox{$e^+p$}}
\newcommand{\emip}{\mbox{$e^-p$}}
\newcommand{\gamm}{$\gamma$}
\newcommand{\zzo}{$Z^{o}$}
%
%
\newcommand{\phib}{\mbox{$\varphi$}}
\newcommand{\rh}{\mbox{$\rho$}}
\newcommand{\rhz}{\mbox{$\rh^0$}}
\newcommand{\ph}{\mbox{$\phi$}}
\newcommand{\om}{\mbox{$\omega$}}
\newcommand{\jpsi}{\mbox{$J/\psi$}}
\newcommand{\pipi}{\mbox{$\pi^+\pi^-$}}
\newcommand{\pip}{\mbox{$\pi^+$}}
\newcommand{\pim}{\mbox{$\pi^-$}}
\newcommand{\kk}{\mbox{K^+K^-$}}
\newcommand{\bsl}{\mbox{$b$}}
\newcommand{\alp}{\mbox{$\alpha^\prime$}}
\newcommand{\alpom}{\mbox{$\alpha_{\PO}$}}
\newcommand{\alregg}{\mbox{$\alpha_{\regg}$}}
\newcommand{\alpomp}{\mbox{$\alpha_{\PO}^\prime$}}
\newcommand{\rzzzz}{\mbox{$r_{00}^{04}$}}
\newcommand{\rzqzz}{\mbox{$r_{00}^{04}$}}
\newcommand{\rzquz}{\mbox{$r_{10}^{04}$}}
\newcommand{\rzqumu}{\mbox{$r_{1-1}^{04}$}}
\newcommand{\ruuu}{\mbox{$r_{11}^{1}$}}
\newcommand{\ruzz}{\mbox{$r_{00}^{1}$}}
\newcommand{\ruuz}{\mbox{$r_{10}^{1}$}}
\newcommand{\ruumu}{\mbox{$r_{1-1}^{1}$}}
\newcommand{\rduz}{\mbox{$r_{10}^{2}$}}
\newcommand{\rdumu}{\mbox{$r_{1-1}^{2}$}}
\newcommand{\rcuu}{\mbox{$r_{11}^{5}$}}
\newcommand{\rczz}{\mbox{$r_{00}^{5}$}}
\newcommand{\rcuz}{\mbox{$r_{10}^{5}$}}
\newcommand{\rcumu}{\mbox{$r_{1-1}^{5}$}}
\newcommand{\rsuz}{\mbox{$r_{10}^{6}$}}
\newcommand{\rsumu}{\mbox{$r_{1-1}^{6}$}}
\newcommand{\rzqik}{\mbox{$r_{ik}^{04}$}}
\newcommand{\rhzik}{\mbox{$\rh_{ik}^{0}$}}
\newcommand{\rhqik}{\mbox{$\rh_{ik}^{4}$}}
\newcommand{\rhaik}{\mbox{$\rh_{ik}^{\alpha}$}}
\newcommand{\rhzzz}{\mbox{$\rh_{00}^{0}$}}
\newcommand{\rhqzz}{\mbox{$\rh_{00}^{4}$}}
\newcommand{\raik}{\mbox{$r_{ik}^{\alpha}$}}
\newcommand{\razz}{\mbox{$r_{00}^{\alpha}$}}
\newcommand{\rauz}{\mbox{$r_{10}^{\alpha}$}}
\newcommand{\raumu}{\mbox{$r_{1-1}^{\alpha}$}}

\newcommand{\R}{\mbox{$R$}}
\newcommand{\rzero}{\mbox{$r_{00}^{04}$}}
\newcommand{\rone}{\mbox{$r_{1-1}^{1}$}}
\newcommand{\costh}{\mbox{$\cos\theta$}}
\newcommand{\cosp}{\mbox{$\cos\psi$}}
\newcommand{\costop}{\mbox{$\cos(2\psi)$}}
\newcommand{\cosd}{\mbox{$\cos\delta$}}
\newcommand{\cossqp}{\mbox{$\cos^2\psi$}}
\newcommand{\cossqt}{\mbox{$\cos^2\theta^*$}}
\newcommand{\sint}{\mbox{$\sin\theta^*$}}
\newcommand{\sintot}{\mbox{$\sin(2\theta^*)$}}
\newcommand{\sinsqt}{\mbox{$\sin^2\theta^*$}}
\newcommand{\costhst}{\mbox{$\cos\theta^*$}}
\newcommand{\vep}{\mbox{$V p$}}
\newcommand{\mpipi}{\mbox{$m_{\pi^+\pi^-}$}}
\newcommand{\mkk}{\mbox{$m_{KK}$}}
\newcommand{\mkaka}{\mbox{$m_{K^+K^-}$}}
\newcommand{\mpp}{\mbox{$m_{\pi\pi}$}}       
\newcommand{\mppsq}{\mbox{$m_{\pi\pi}^2$}}   
\newcommand{\mpi}{\mbox{$m_{\pi}$}}          
\newcommand{\mrho}{\mbox{$m_{\rho}$}}        
\newcommand{\mrhosq}{\mbox{$m_{\rho}^2$}}    
\newcommand{\Gmpp}{\mbox{$\Gamma (\mpp)$}}   
\newcommand{\Gmppsq}{\mbox{$\Gamma^2(\mpp)$}}
\newcommand{\Grho}{\mbox{$\Gamma_{\rho}$}}   
\newcommand{\grho}{\mbox{$\Gamma_{\rho}$}}   
\newcommand{\Grhosq}{\mbox{$\Gamma_{\rho}^2$}}   
%
%
\newcommand{\cm}{\mbox{\rm cm}}
\newcommand{\GeV}{\mbox{\rm GeV}}
\newcommand{\gev}{\mbox{\rm GeV}}
\newcommand{\GeVx}{\rm GeV}
\newcommand{\gevx}{\rm GeV}
\newcommand{\GeVc}{\rm GeV/c}
\newcommand{\gevc}{\rm GeV/c}
\newcommand{\MeVc}{\rm MeV/c}
\newcommand{\mevc}{\rm MeV/c}
\newcommand{\MeV}{\mbox{\rm MeV}}
\newcommand{\mev}{\mbox{\rm MeV}}
\newcommand{\MeVx}{\mbox{\rm MeV}}
\newcommand{\mevx}{\mbox{\rm MeV}}
\newcommand{\GeVsq}{\mbox{${\rm GeV}^2$}}
\newcommand{\gevsq}{\mbox{${\rm GeV}^2$}}
\newcommand{\gevsqc}{\mbox{${\rm GeV^2/c^4}$}}
\newcommand{\gevcsq}{\mbox{${\rm GeV/c^2}$}}
\newcommand{\mevcsq}{\mbox{${\rm MeV/c^2}$}}
\newcommand{\GeVsqm}{\mbox{${\rm GeV}^{-2}$}}
\newcommand{\gevsqm}{\mbox{${\rm GeV}^{-2}$}}
\newcommand{\nb}{\mbox{${\rm nb}$}}
\newcommand{\nbinv}{\mbox{${\rm nb^{-1}}$}}
\newcommand{\pbinv}{\mbox{${\rm pb^{-1}}$}}
\newcommand{\mm}{\mbox{$\cdot 10^{-2}$}}
\newcommand{\mmm}{\mbox{$\cdot 10^{-3}$}}
\newcommand{\mmmm}{\mbox{$\cdot 10^{-4}$}}
\newcommand{\degr}{\mbox{$^{\circ}$}}
%
%
\newcommand{\F}{$ F_{2}(x,Q^2)\,$}  
\newcommand{\Fc}{$ F_{2}\,$}    
\newcommand{\XP}{x_{{I\!\!P}/{p}}}       
\newcommand{\TOSS}{x_{{i}/{\PO}}}        
\newcommand{\un}[1]{\mbox{\rm #1}} 
\newcommand{\LO}{Leading Order}
\newcommand{\NLO}{Next to Leading Order}
\newcommand{\ft}{$ F_{2}\,$}
%
%
\newcommand{\mc}{\multicolumn}
\newcommand{\bce}{\begin{center}}
\newcommand{\ece}{\end{center}}
\newcommand{\beq}{\begin{equation}}
\newcommand{\eeq}{\end{equation}}
\newcommand{\bea}{\begin{eqnarray}}
\newcommand{\eea}{\end{eqnarray}}
%
%
\def\lsim{\mathrel{\rlap{\lower4pt\hbox{\hskip1pt$\sim$}}
    \raise1pt\hbox{$<$}}}         
\def\gsim{\mathrel{\rlap{\lower4pt\hbox{\hskip1pt$\sim$}}
    \raise1pt\hbox{$>$}}}         
%
%
\newcommand{\pom}{{I\!\!P}}
\newcommand{\regg}{{I\!\!R}}
\newcommand{\PO}{I\!\!P}
\newcommand{\slowpi}{\pi_{\mathit{slow}}}
\newcommand{\fiidiii}{F_2^{D(3)}}
\newcommand{\fiidiiiarg}{F_2^{D(3)}\,(\beta,\,Q^2,\,x)}
\newcommand{\fiidiiifull}{F_2^{D(3)}\,(x_{I\!\!P},\,\beta,\,Q^2)}
\newcommand{\n}{1.19\pm 0.06 (stat.) \pm0.07 (syst.)}
\newcommand{\nz}{1.30\pm 0.08 (stat.)^{+0.08}_{-0.14} (syst.)}
\newcommand{\fiidiiiifull}{$F_2^{D(4)}\,(\beta,\,Q^2,\,x,\,t)$}
\newcommand{\fiipom}{\tilde F_2^D}
\newcommand{\fiipomfull}{\tilde F_2^D\,(\beta,\,Q^2)}
\newcommand{\ALPHA}{1.10\pm0.03 (stat.) \pm0.04 (syst.)}
\newcommand{\ALPHAZ}{1.15\pm0.04 (stat.)^{+0.04}_{-0.07} (syst.)}
\newcommand{\fiipomarg}{\fiipom\,(\beta,\,Q^2)}
\newcommand{\pomflux}{f_{\pom / p}}
\newcommand{\nxpom}{1.19\pm 0.06 (stat.) \pm0.07 (syst.)}
\newcommand {\gapprox}
   {\raisebox{-0.7ex}{$\stackrel {\textstyle>}{\sim}$}}
\newcommand {\lapprox}
   {\raisebox{-0.7ex}{$\stackrel {\textstyle<}{\sim}$}}
\newcommand{\pomfluxarg}{f_{\pom / p}\,(x_\pom)}
\newcommand{\dsf}{\mbox{$F_2^{D(3)}$}}
\newcommand{\dsfva}{\mbox{$F_2^{D(3)}(\beta,Q^2,x_{I\!\!P})$}}
\newcommand{\dsfvb}{\mbox{$F_2^{D(3)}(\beta,Q^2,x)$}}
\newcommand{\dsfpom}{$F_2^{I\!\!P}$}
\newcommand{\gap}{\stackrel{>}{\sim}}
\newcommand{\lap}{\stackrel{<}{\sim}}
\newcommand{\fem}{$F_2^{em}$}
\newcommand{\tsnmp}{$\tilde{\sigma}_{NC}(e^{\mp})$}
\newcommand{\tsnm}{$\tilde{\sigma}_{NC}(e^-)$}
\newcommand{\tsnp}{$\tilde{\sigma}_{NC}(e^+)$}
\newcommand{\st}{$\star$}
\newcommand{\sst}{$\star \star$}
\newcommand{\ssst}{$\star \star \star$}
\newcommand{\sssst}{$\star \star \star \star$}
\newcommand{\tw}{\theta_W}
\newcommand{\sw}{\sin{\theta_W}}
\newcommand{\cw}{\cos{\theta_W}}
\newcommand{\sww}{\sin^2{\theta_W}}
\newcommand{\cww}{\cos^2{\theta_W}}
\newcommand{\trm}{m_{\perp}}
\newcommand{\trp}{p_{\perp}}
\newcommand{\trmm}{m_{\perp}^2}
\newcommand{\trpp}{p_{\perp}^2}
%
%
\newcommand{\sqrts}{$\sqrt{s}$}
\newcommand{\Oa}{$O(\alpha_s)$}
\newcommand{\Oaa}{$O(\alpha_s^2)$}
\newcommand{\PT}{p_{\perp}}
\newcommand{\sh}{\hat{s}}
\newcommand{\uh}{\hat{u}}
\newcommand{\ttbs}{\char'134}
\newcommand{\xpomlo}{3\times10^{-4}}
\newcommand{\xpomup}{0.05}
\newcommand{\llq}{$\alpha_s \ln{(\qsq / \Lambda_{QCD}^2)}$}
\newcommand{\llqx}{$\alpha_s \ln{(\qsq / \Lambda_{QCD}^2)} \ln{(1/x)}$}
\newcommand{\llx}{$\alpha_s \ln{(1/x)}$}
%
%
%
%
\def\ar#1#2#3   {{\em Ann. Rev. Nucl. Part. Sci.} {\bf#1} (#2) #3}
\def\epj#1#2#3  {{\em Eur. Phys. J.} {\bf#1} (#2) #3}
\def\err#1#2#3  {{\it Erratum} {\bf#1} (#2) #3}
\def\ib#1#2#3   {{\it ibid.} {\bf#1} (#2) #3}
\def\ijmp#1#2#3 {{\em Int. J. Mod. Phys.} {\bf#1} (#2) #3}
\def\jetp#1#2#3 {{\em JETP Lett.} {\bf#1} (#2) #3}
\def\mpl#1#2#3  {{\em Mod. Phys. Lett.} {\bf#1} (#2) #3}
\def\nim#1#2#3  {{\em Nucl. Instr. Meth.} {\bf#1} (#2) #3}
\def\nc#1#2#3   {{\em Nuovo Cim.} {\bf#1} (#2) #3}
\def\np#1#2#3   {{\em Nucl. Phys.} {\bf#1} (#2) #3}
\def\pl#1#2#3   {{\em Phys. Lett.} {\bf#1} (#2) #3}
\def\prep#1#2#3 {{\em Phys. Rep.} {\bf#1} (#2) #3}
\def\prev#1#2#3 {{\em Phys. Rev.} {\bf#1} (#2) #3}
\def\prl#1#2#3  {{\em Phys. Rev. Lett.} {\bf#1} (#2) #3}
\def\ptp#1#2#3  {{\em Prog. Th. Phys.} {\bf#1} (#2) #3}
\def\rmp#1#2#3  {{\em Rev. Mod. Phys.} {\bf#1} (#2) #3}
\def\rpp#1#2#3  {{\em Rep. Prog. Phys.} {\bf#1} (#2) #3}
\def\sjnp#1#2#3 {{\em Sov. J. Nucl. Phys.} {\bf#1} (#2) #3}
\def\spj#1#2#3  {{\em Sov. Phys. JEPT} {\bf#1} (#2) #3}
\def\zp#1#2#3   {{\em Zeit. Phys.} {\bf#1} (#2) #3}
%
%
\newcommand{\clearemptydoublepage}{\newpage{\pagestyle{empty}\cleardoublepage}}
\newcommand{\scaption}[1]{\caption{\protect{\footnotesize  #1}}}
\newcommand{\proc}[2]{\mbox{$ #1 \rightarrow #2 $}}
\newcommand{\average}[1]{\mbox{$ \langle #1 \rangle $}}
\newcommand{\av}[1]{\mbox{$ \langle #1 \rangle $}}

%% file: hel.bbl
\section*{References}
\begin{thebibliography}{99}
%
\bibitem{lambda}
H1 Coll., 
S. Aid et al., 
\np {B470} {1996} {3}.

%
\bibitem{DoLa}
see e.g. A. Donnachie, P.V. Landshoff,
\pl {B296}{1992} {227}.
%
\bibitem{H194}
H1 Coll., 
C. Adloff et al., 
\zp {C76} {1997} {613}.
%
\bibitem{ZEUSMX}
ZEUS Coll., J. Breitweg et al.,
\epj {C6} {1999}{43}.
%
\bibitem{ZEUSBPC}
ZEUS Coll.,
paper 875, ICHEP 2000, Osaka, Japan.
%
\bibitem{f2d-ZEUSLPS}
ZEUS Coll.,
J. Breitweg et al.,
\epj {C1} {1998}{61}. 
%
\bibitem{IS}
G.~Ingelman and P.~Schlein, \pl {B152} {1985} {256}.
%
\bibitem{collins}
J.C. Collins,
\prev {D57} {1998} {3051}; {\bf D61} (2000) 019902.
%
\bibitem{tslope-ZEUSLPS}
ZEUS Coll.,
J. Breitweg et al.,
\epj {C2} {1998}{237}. 
%
\bibitem{paul}
P.~R.~Newman,
ICHEP 2000, Osaka, Japan.
%
\bibitem{photoprod-h1}
H1 Coll., 
C. Adloff et al., 
\zp {C74} {1997} {221}.
%
\bibitem{photoprod-zeus}
ZEUS Coll.,
J. Breitweg et al.,
\zp {C75} {1997} {421}.
%
\bibitem{amirim}
for a review, see e.g. 
M.~F.~McDermott, hep-ph/0008260 (2000).
%
\bibitem{rho-photoprod-h1}
H1 Coll., 
S. Aid et al., 
\np {b463} {1996} {3}.
%
\bibitem{rho-photoprod-zeus}
ZEUS Coll.,
J. Breitweg et al.,
\epj {C2} {1998} {247}.
%
\bibitem{psi-h1}
H1 Coll., 
C. Adloff et al., 
\pl {B483} {2000} {23}.
%
\bibitem{psi-zeus}
ZEUS Coll.,
paper 878, ICHEP 2000, Osaka, Japan.
%
\bibitem{bekw}
J. Bartels et al.,
\epj {C7} {1999} {443}.
%
\bibitem{cdf}
CDF Coll.,
T. Affolder et al., 
\prl {84} {2000} {232}.
%
\bibitem{zeus-final-state}
ZEUS Coll.,
J. Breitweg et al.,
\pl {B421} {1998} {368}; 
paper 876, ICHEP 2000, Osaka, Japan.
%
\bibitem{h1-final-state}
H1 Coll., 
C. Adloff et al., 
\epj {C1} {1998} {495};
\pl {B428} {1998} {206};
\epj {C5} {1998} {439}.
%
\bibitem{h1-jets}
H1 Coll., 
C. Adloff et al., 
hep-ex/0012051.
%
\bibitem{zeus-jets}
ZEUS Coll.,
J. Breitweg et al.,
\epj {C5} {1998} {41}.
%
\bibitem{zeus-charm}
ZEUS Coll.,
paper 874, ICHEP 2000, Osaka, Japan.
%
\bibitem{h1-charm}
H1 Coll.,
paper 157ag, HEP99, Tampere, Finland.
%
\bibitem{rho-zeus-osaka}
ZEUS Coll.,
paper 880, ICHEP 2000, Osaka, Japan.
%
\bibitem{rho-h1}
H1 Coll., 
C. Adloff et al., 
\epj {C13} {2000} {371}.
%
\bibitem{phi-h1}
H1 Coll., 
C. Adloff et al., 
\pl {B483} {2000} {360}.
%
\bibitem{yoshida}
R. Yoshida, 
RADCOR 2000, Carmel, USA.
%
\bibitem{rho-zeus-mde}
ZEUS Coll.,
J. Breitweg et al.,
\epj {C12} {2000} {393}.
%
\bibitem{ivanov}
D.~Yu.~Ivanov and R.~Kirschner,
\prev {D58} {1998} {114026}.
%
\bibitem{dvcs-h1}
H1 Coll., 
paper 966, ICHEP 2000, Osaka, Japan.
%
\bibitem{spd}
see e.g. A.~V.~Radyushkin,
\prev {D56} {1997} {5524}.
%
\bibitem{zeus-upsilon}
ZEUS Coll.,
J. Breitweg et al.,
\pl {B487} {1998} {432}.
%
\bibitem{pots}
http://web.iihe.ac.be/h1pot/
%

\end{thebibliography}
